\documentclass[a4paper]{elsart}
\usepackage{graphicx}
\usepackage{times}

\begin{document}

\journal{Physics Letters B}

\begin{frontmatter}

\title{First measurement of the $\pi^+\pi^-$ atom lifetime }


\author[p]{B.~Adeva},
\author[l]{L.~Afanasyev\corauthref{cor}},
\corauth[cor]{PH Division, CERN, CH 1211 Geneva 23, Switzerland;\\ 
e-mail: Leonid.Afanasev@cern.ch}
\author[d]{M.~Benayoun},
\author[q]{A.~Benelli},
\author[b]{Z.~Berka},
\author[o]{V.~Brekhovskikh},
\author[m]{G.~Caragheorgheopol},
\author[b]{T.~Cechak},
\author[j]{M.~Chiba},
\author[m]{S.~Constantinescu},
\author[a]{C.~Detraz},
\author[f]{D.~Dreossi},
\author[a]{D.~Drijard},
\author[l]{A.~Dudarev},
\author[s]{I.~Evangelou},
\author[a]{M.~Ferro-Luzzi},
\author[p,a]{M.V.~Gallas},
\author[b]{J.~Gerndt},
\author[f]{R.~Giacomich},
\author[e]{P.~Gianotti},
\author[q]{D.~Goldin},
\author[p]{F.~G\'omez},
\author[o]{A.~Gorin},
\author[l]{O.~Gorchakov},
\author[e]{C.~Guaraldo},
\author[a]{M.~Hansroul},
\author[b]{R.~Hosek},
\author[e,m]{M.~Iliescu},
\author[l]{V.~Karpukhin},
\author[b]{J.~Kluson},
\author[g]{M.~Kobayashi},
\author[s]{P.~Kokkas},
\author[l]{V.~Komarov},
\author[l]{V.~Kruglov},
\author[l]{L.~Kruglova},
\author[l]{A.~Kulikov},
\author[l]{A.~Kuptsov},
\author[o]{I.~Kurochkin},
\author[l]{K.-I.~Kuroda},
\author[f]{A.~Lamberto},
\author[a,e]{A.~Lanaro},
\author[o]{V.~Lapshin},
\author[c]{R.~Lednicky},
\author[d]{P.~Leruste},
\author[e]{P.~Levi Sandri},
\author[p]{A.~Lopez Aguera},
\author[e]{V.~Lucherini},
\author[i]{T.~Maki},
\author[s]{N.~Manthos},
\author[o]{I.~Manuilov},
\author[a]{L.~Montanet},
\author[d]{J.-L.~Narjoux},
\author[a,l]{L.~Nemenov},
\author[l]{M.~Nikitin},
\author[p]{T.~N\'u\~nez Pardo},
\author[h]{K.~Okada},
\author[l]{V.~Olchevskii},
\author[p]{A.~Pazos},
\author[m]{M.~Pentia},
\author[f]{A.~Penzo},
\author[a]{J.-M.~Perreau},
\author[e,m]{C.~Petrascu},
\author[p]{M.~Pl\'o},
\author[m]{T.~Ponta},
\author[m]{D.~Pop},
\author[f]{G.F.~Rappazzo},
\author[p]{A.~Rodriguez Fernandez},
\author[p]{A.~Romero},
\author[o]{A.~Ryazantsev},
\author[o]{V.~Rykalin},
\author[p,q,a]{C.~Santamarina},
\author[p]{J.~Saborido},
\author[r]{J.~Schacher},
\author[q]{Ch.P.~Schuetz},
\author[o]{A.~Sidorov},
\author[c]{J.~Smolik},
\author[h]{F.~Takeutchi},
\author[l]{A.~Tarasov},
\author[q]{L.~Tauscher},
\author[p]{M.J.~Tobar},
\author[n]{S.~Trusov},
\author[l]{V.~Utkin},
\author[p]{O.~V\'azquez Doce},
\author[p]{P.~V\'azquez},
\author[q]{S.~Vlachos},
\author[n]{V.~Yazkov},
\author[g]{Y.~Yoshimura},
\author[l]{M.~Zhabitsky},
\author[l]{P.~Zrelov}

\address[a]{CERN, Geneva, Switzerland}
\address[b]{Czech Technical University, Prague, Czech Republic}
\address[c]{Institute of Physics ACSR, Prague, Czech Republic}
\address[s]{Ioannina University, Ioannina, Greece}
\address[d]{LPNHE des Universites Paris VI/VII, IN2P3-CNRS, France}
\address[e]{INFN - Laboratori Nazionali di Frascati, Frascati, Italy}
\address[f]{INFN - Trieste and Trieste University, Trieste, Italy}
\address[g]{KEK, Tsukuba, Japan}
\address[h]{Kyoto Sangyo University, Kyoto, Japan}
\address[i]{UOEH-Kyushu, Japan}
\address[j]{ Tokyo Metropolitan University, Japan}
\address[l]{JINR Dubna, Russia}
\address[m]{IFIN-HH, National Institute for Physics and Nuclear Engineering, Bucharest, Romania}
\address[n]{Skobeltsin Institute for Nuclear Physics of Moscow State University Moscow, Russia}
\address[o]{IHEP Protvino, Russia}
\address[p]{Santiago de Compostela University, Spain}
\address[q]{Basel University, Switzerland}
\address[r]{Bern University, Switzerland}

\begin{abstract}
  The goal of the DIRAC experiment at CERN (PS212) is to measure the
  $\pi^+\pi^-$~atom lifetime with $10\%$ precision. Such a measurement
  would yield a precision of $5\%$ on the value of the $S$-wave
  $\pi\pi$ scattering lengths combination $\left|a_0-a_2\right|$.
  Based on part of the collected data we present a first result on the
  lifetime, $\tau=\left[2.91~^{+0.49}_{-0.62}\right]\times 10^{-15}$
  s, and discuss the major systematic errors. This lifetime
  corresponds to
  $\left|a_0-a_2\right|=0.264~^{+0.033}_{-0.020}~m_{\pi}^{-1}$.

\end{abstract}


\end{frontmatter}


\section{Introduction}
\label{sec:intro}

The aim of the DIRAC experiment at CERN \cite{Adeva95} is to measure
the lifetime of pionium, an atom consisting of a $\pi^+$ and a $\pi^-$
meson ($A_{2\pi}$).  The lifetime is dominated by the charge-exchange
scattering process ($\pi^+ \pi^{-} \rightarrow \pi^{0} \pi^{0}$)
\footnote{Annihilation into two photons amounts to $\approx$ 0.3\%
  \cite{Uretsky61,hammer} and is neglected here.}  and is thus related
to the relevant scattering lengths \cite{Deser54}.  The partial decay
width of the atomic ground state (principal quantum number $n=1$,
orbital quantum number $l=0$) is
\cite{Uretsky61,Bilenky69,Jallouli98,Ivanov98,Gasser01,Gashi02}

\begin{equation}
  \label{eq:gasser}
   \Gamma_{1S}=\frac{1}{\tau_{1S}}=
   \frac{2}{9} \; \alpha^3 \; p\left|a_0-a_2\right|^2 (1+\delta)\
\end{equation}

with $\tau_{1S}$ the lifetime of the atomic ground state, $\alpha$ the
fine-structure constant, $p$ the $\pi^0$ momentum in the atomic rest
frame, and $a_0$ and $a_2$ the $S$-wave $\pi\pi$ scattering lengths
for isospin 0 and 2, respectively.  The term $\delta$ accounts for QED
and QCD corrections \cite{Jallouli98,Ivanov98,Gasser01,Gashi02}. It is
a known quantity ($\delta=(5.8\pm1.2)\times10^{-2}$) ensuring a 1\%
accuracy for Eq.~(\ref{eq:gasser}) \cite{Gasser01}.  A measurement of
the lifetime therefore allows to obtain in a model-independent way
the value of $|a_0-a_2|$.  The $\pi\pi$ scattering lengths $a_0$,
$a_2$ have been calculated within the framework of Standard Chiral
Perturbation Theory \cite{Weinb79} with a precision better than 2.5\%
\cite{Colan01NP} ($a_{0}=0.220\pm 0.005$, $a_{2}=-0.0444\pm0.0010$,
$a_0-a_2=0.265\pm0.004$ in units of inverse pion mass) and lead to the
prediction $\tau_{1S}=(2.9\pm0.1)\times10^{-15}$~s. The Generalized
Chiral Perturbation Theory though allows for larger $a$-values
\cite{Knecht95}.  Model independent measurements of $a_{0}$ have been
done using $K_{e4}$ decays \cite{Ross77,Pislak01}.

Oppositely charged pions emerging from a high energy proton-nucleus
collision may be either produced directly or stem from strong decays
("short-lived" sources) and electromagnetic or weak decays
("long-lived'' sources) of intermediate hadrons.  Pion pairs from
``short-lived'' sources undergo Coulomb final state interaction and
may form atoms. The region of production being small as compared to
the Bohr radius of the atom and neglecting strong final state
interaction, the cross section $\sigma^{n}_{A}$ for production of
atoms with principal quantum number $n$ is related to the inclusive
production cross section for pion pairs from "short lived" sources
without Coulomb correlation ($\sigma^0_s$) \cite{Nem85} :

\begin{equation}
   \label{eq:atomprod}
   \frac{d\sigma^n_A}{d{\vec{p}}_A}={(2\pi)}^3\frac{E_{A}}{M_{A}}|\Psi_n^{C}(\vec{r}^{*}=0)|^2
   \left.\frac{d^{2}\sigma^0_s}{d{\vec{p}}_+ d{\vec{p}}_-}\right|_{\vec{p}_+=
   \vec{p}_-}
\end{equation}

with $\vec{p}_A$, $E_A$ and $M_{A}$ the momentum, energy and mass of
the atom in the lab frame, respectively, and $\vec{p}_+$, $\vec{p}_-$
the momenta of the charged pions. The square of the Coulomb atomic
wave function for zero distance $\vec{r}^{*}$ between them in the c.m.
system is $| \Psi_n^{C}(0) |^2=p^3_B/\pi n^3$, where $p_B=m_\pi
\alpha/2$ is the Bohr momentum of the pions and $m_{\pi}$ the pion
mass. The production of atoms occurs only in $S$-states \cite{Nem85}.

Final state interaction also transforms the ``unphysical'' cross
section $\sigma^0_s$ into a real one for Coulomb correlated pairs,
$\sigma_{C}$   \cite{Sakh48,lednicki}:

\begin{equation}\label{eq:coulomb}
    \frac{d^2\sigma_{C}}{d\vec{p}_+d\vec{p}_-}=|\Psi_{-\vec{k}^{*}}^{C}(\vec{r}^{*})|^{2}
    \frac{d^2\sigma^0_s}{d\vec{p}_+d\vec{p}_-},
\end{equation}

where $\Psi_{-\vec{k}^{*}}^{C}(\vec{r}^{*})$ is the continuum wave
function and $2\vec{k}^{*}\equiv \vec{q}$ with $\vec{q}$ being the
relative momentum of the $\pi^+$ and $\pi^-$ in the c.m.
system\footnote{For the sake of clarity we use the symbol $Q$ for the
experimentally reconstructed and $q$ for the physical relative
momentum.}.  $|\Psi_{-\vec{k}^{*}}^{C}(\vec{r}^{*})|^{2}$ describes
the Coulomb correlation and at $r^{*}=0$ coincides with the
Gamov-Sommerfeld factor $A_C(q)$ with $q=|\vec{q}|$ \cite{lednicki}:

\begin{equation}
A_C(q)=
\frac{2\pi m_{\pi}\alpha/q}{1-\exp(-2\pi m_{\pi}\alpha/q)}\;.
\label{eq:ac}
\end{equation}

For low $q$, $0\leq q\leq q_{0}$, Eqs.~(\ref{eq:atomprod},
\ref{eq:coulomb}, \ref{eq:ac}) relate the number of produced
$A_{2\pi}$ atoms, $N_{A}$, to the number of Coulomb correlated pion
pairs, $N_{CC}$ \cite{AFAN97} :

\begin{equation}
\frac{N_{A}}{N_{CC}}=
\frac{\sigma_A^{tot}}{\sigma_{C}^{tot}|_{q\leq q_{0}}}
=\frac{(2\pi \alpha m_{\pi})^{3}}{\pi}
\frac{\displaystyle{\sum\limits_{n=1}^{\infty}\frac{1}{n^3} }}
{\displaystyle{\int\limits_0^{q_{0}}A_C(q)d^{3}q}}
=k_{th}(q_0).
\label{eq:kfactor}
\end{equation}

Eq.~(\ref{eq:kfactor}) defines the theoretical $k$-factor. Throughout 
the paper we will use 
\begin{equation}
    q_{0} = 2 \mathrm{MeV}/c ~\mathrm{~and} ~k_{th}(q_{0})=0.615.
    \label{eq:kth}
\end{equation}

In order to account for the finite size of the pion production region
and of the two-pion final state strong interaction, the squares of the
Coulomb wave functions in Eqs.~(\ref{eq:atomprod}) and
(\ref{eq:coulomb}) must be substituted by the square of the complete
wave functions, averaged over the distance $\vec{r}^{*}$ and the
additional contributions from $\pi^{0}\pi^{0}\rightarrow A_{2\pi}$ as
well as $\pi^{0}\pi^{0}\rightarrow \pi^{+}\pi^{-}$ \cite{lednicki}.
It should be noticed that these corrections essentially cancel in the
$k$-factor (Eq. (\ref{eq:kfactor})) and lead to a correction of only a
fraction of a percent. Thus finite size corrections can safely be
neglected for $k_{th}$.

\begin{figure}[htb]
  \centering \includegraphics[width=0.47\textwidth]{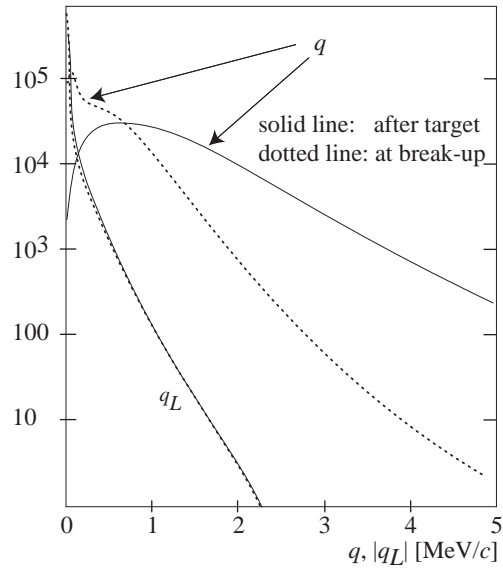}
\caption{ Relative momentum distributions ($q$, $q_{L}$) for atomic $\pi^{+}\pi^{-}$ pairs 
at the point of break-up and at the exit of the target. 
Note that $q_{L}$ is almost not affected by multiple scattering in the target.} 
\label{fig:At}
\end{figure}

Once produced, the $A_{2\pi}$ atoms propagate with relativistic
velocity (average Lorentz factor $\bar{\gamma} \approx 17$ in our
case) and, before they decay, interact with target atoms, whereby they
become excited/de-excited or break up. The $\pi^{+}\pi^{-}$ pairs from
break-up (atomic pairs) exhibit specific kinematical features which
allow to identify them experimentally \cite{Nem85}, namely very low
relative momentum $q$ and $q_{L}$ (the component of $\vec{q}$ parallel
to the total momentum $\vec p_{+}+\vec p_{-}$) as shown in Fig.
\ref{fig:At}. After break-up, the atomic pair traverses the target and
to some extent loses these features by multiple scattering,
essentially in the transverse direction, while $q_{L}$ is almost not
affected. This is one reason for considering distributions in $Q_{L}$
as well as in $Q$ when analyzing the data.

Excitation/de-excitation and break-up of the atom are competing with
its decay. Solving the transport equations with the cross sections for
excitation and break-up,
\cite{Dulian83,Mrowc,Afan96,Halab99,Taras91,Voskr98,Afan99,Ivanov99gl,Heim00,Heim01,Schum02,Afan02}
leads to a target-specific relation between break-up probability and
lifetime which is estimated to be accurate at the 1\% level
\cite{Afan96,santa03,afan04}.  Measuring the break-up probability thus
allows to determine the lifetime of pionium \cite{Nem85}.

The first observation of the $A_{2\pi}$ atom \cite{Afan93} has allowed
to set a lower limit on its lifetime \cite{AFAN97,Afan94} of $\tau >
1.8\times10^{-15}$~s (90\%~CL).  In this paper we present a
determination of the lifetime of the $A_{2\pi}$ atom, based on a large
sample of data taken in 2001 with Ni targets.


\section{The DIRAC experiment}

The DIRAC experiment uses a magnetic double-arm spectrometer at the
CERN 24~GeV/$c$ extracted proton beam T8. Details on the set-up may be
found in \cite{Setup03}. Since its start-up, DIRAC has accumulated
about 15'000 atomic pairs. The data used for this work were taken with
two Ni foils, one of 94$\mu$m thickness (76\% of the $\pi^{+}\pi^{-}$
data), and one of 98 $\mu$m thickness (24\% of the data). An extensive
description of the DIRAC set-up, data selection, tracking, Monte Carlo
procedures, signal extraction and a first high statistics
demonstration of the feasibility of the lifetime measurement, based on
the Ni data of 2001, have been published in \cite{signalpaper}.

\begin{figure}[ht]
\begin{center}
\includegraphics[width=0.8\textwidth]{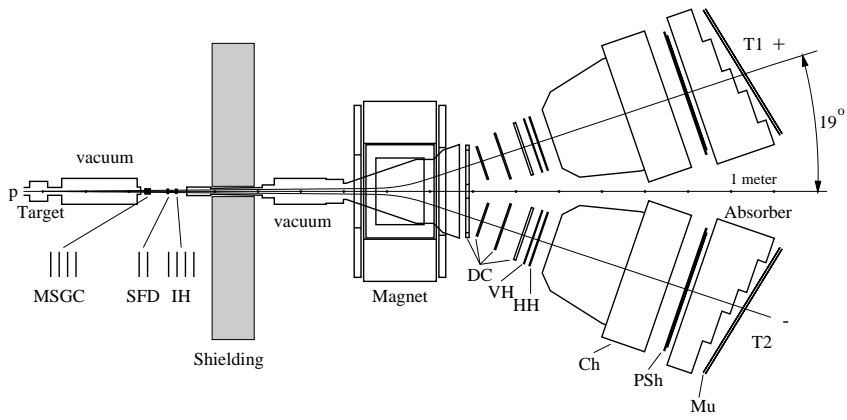}
\caption{Schematic top view of the DIRAC spectrometer. Upstream of the
magnet: target, microstrip gas chambers (MSGC), scintillating fiber
detectors (SFD), ionization hodoscopes (IH) and iron shielding. Downstream
of the magnet: drift chambers (DC), vertical and horizontal scintillation
hodoscopes (VH, HH), gas Cherenkov counters (Ch), preshower detectors (PSh)
and, behind the iron absorber, muon detectors (Mu).}
\label{fig:schem}
\end{center}
\end{figure}

The set-up and the definitions of detector acronyms are shown in Fig.
\ref{fig:schem}. The main selection criteria and
performance parameters \cite{signalpaper} are recalled in the following.

Pairs of oppositely charged pions are selected by means of Cherenkov,
preshower and muon counters. Through the measurement of the time
difference between the vertical hodoscope signals of the two arms,
time correlated (prompt) events ($\sigma_{\Delta t}=185$\,ps) can be
distinguished from accidental events (see \cite{signalpaper}).  The
resolution of the three components of the relative momentum $Q$ of two
tracks, transverse and parallel to the c.m. flight direction, $Q_{x}$,
$Q_{y}$ and $Q_{L}$, is about $0.5~\mathrm{MeV}/c$ for $Q\le
4$\,MeV/$c$.  Due to charge combinatorials and inefficiencies of the
SFD, the distributions for the transverse components have substantial
tails, which the longitudinal component does not exhibit
\cite{schuetzthesis}. This is yet another reason for analyzing both
$Q$ and $Q_{L}$ distributions.

Data were analyzed with the help of the DIRAC analysis software
package ARIANE \cite{ariane}.

The tracking procedures require the two tracks either to have
a common vertex in the target plane (``V-tracking'') or to originate
from the intersect of the beam with the target (``T-tracking'').  In
the following we limit ourselves to quoting results obtained with
T-tracking.  Results obtained with V-tracking do not show significant
differences, as will be shown later.

The following cuts and conditions are applied (see \cite{signalpaper}):

\begin{itemize}
    
\item at least one track candidate per arm with a confidence level
  better than 1\% and a distance to the beam spot in the target
  smaller than 1.5\,cm in x and y;

\item ``prompt'' events are defined by the time difference of the
  vertical hodoscopes in the two arms of the spectrometer of $|\Delta
  t |\leq 0.5$\,ns;

\item ``accidental'' events are defined by time intervals
  $-15\,\mathrm{ns} \leq \Delta t \leq -5$\,ns and
  $7\,\mathrm{ns}\leq\Delta t \leq 17$\,ns, determined by the
  read-out features of the SFD detector (time dependent merging of
  adjacent hits) and exclusion of correlated $\pi^-$p pairs.
  \cite{signalpaper};
  
\item protons in ``prompt'' events are rejected by time-of-flight in
  the vertical hodoscopes for momenta of the positive particle below
  4~GeV/$c$.  Positive particles with higher momenta are rejected;

\item $e^{\pm}$ and $\mu^{\pm}$ are rejected by appropriate cuts
  on the Cherenkov, the preshower and the muon counter information;
  
\item cuts in the transverse and longitudinal components of $Q$ are
  $Q_{T} \leq 4$\,MeV/$c$ and $|Q_L|< 15$\,MeV/$c$.  The $Q_{T}$ cut
  preserves 98\% of the atomic signal. The $Q_{L}$ cut preserves data
  outside the signal region for defining the background;
  

\item only events with at most two preselected hits per SFD plane are
  accepted. This provides the cleanest possible event pattern. 

\end{itemize}


\section{Analysis}

The spectrometer including the target is fully simulated by GEANT-DIRAC
\cite{geant}, a GEANT3-based simulation code. The detectors, including
read-out, inefficiency, noise and digitalization are simulated and
implemented in the DIRAC analysis code ARIANE \cite{ariane}. The
triggers are fully simulated as well.

The simulated data sets for different event types can therefore be
reconstructed with exactly the same procedures
and cuts as used for experimental data.

The different event types are generated according to the underlying physics.

\textbf{Atomic pairs:} Atoms are generated according to Eq.
(\ref{eq:atomprod}) using measured total momentum distributions for
short-lived pairs. The atomic $\pi^{+}\pi^{-}$ pairs are generated
according to the probabilities and kinematics described by the
evolution of the atom while propagating through the target and by the
break-up process (see \cite{santa2003-9}). These $\pi^{+}\pi^{-}$
pairs, starting from their spatial production point, are then
propagated through the remaining part of the target and the full
spectrometer using GEANT-DIRAC. Reconstruction of the track pairs
using the fully simulated detectors and triggers leads to the atomic
pair distribution $dn_{A}^{MC}/dQ$.

\textbf{Coulomb correlated $\pi^{+}\pi^{-}$ pairs (CC-background):} The events are 
generated according to Eqs.~(\ref{eq:coulomb},\ref{eq:ac}) using measured
total momentum distributions for short-lived pairs. The generated $q$-distributions
are assumed to follow phase space modified by the Coulomb correlation 
function (Eq. (\ref{eq:ac})), $dN_{CC}^{gen}/dq\propto q^{2}\times A_{C}(q)$.
Processing them with GEANT-DIRAC and then analyzing them using the full
detector and trigger simulation leads to the Coulomb correlated distribution $dN_{CC}^{MC}/dQ$.

\textbf{Non-correlated $\pi^{+}\pi^{-}$ pairs (NC-background):} $\pi^{+}\pi^{-}$ pairs,
where at least one pion originates from the decay of a "long-lived"
source (e.g. electromagnetically or weakly decaying mesons or baryons)
do not undergo any final state interactions. Thus they are generated 
according to $dN_{NC}^{gen}/dq\propto q^{2}$, using slightly softer
momentum distributions than for short-lived sources (difference obtained from
FRITIOF-6). The Monte Carlo distribution $dN_{NC}^{MC}/dQ$ is obtained as
above.

\textbf{Accidental $\pi^{+}\pi^{-}$ pairs (acc-background):} $\pi^{+}\pi^{-}$ pairs,
where the two pions originate from two different proton-nucleus
interactions, are generated 
according to $dN_{acc}^{gen}/dq\propto q^{2}$, using measured
momentum distributions. The Monte Carlo distribution $dN_{acc}^{MC}/dQ$ 
is obtained as above.

All the Monte Carlo distributions are normalized,
$\int_{0}^{Q_{max}}(dN_{i}^{MC}/dQ) dQ =N_{i}^{MC},~i=CC,NC,acc$, with
statistics about 5 to 10 times higher than the experimental data;
similarly for atomic pairs ($n^{MC}_{A}$).

\begin{figure}[htb]
 \centering 
$\begin{array}{cc}
   \includegraphics[width=0.47\textwidth]{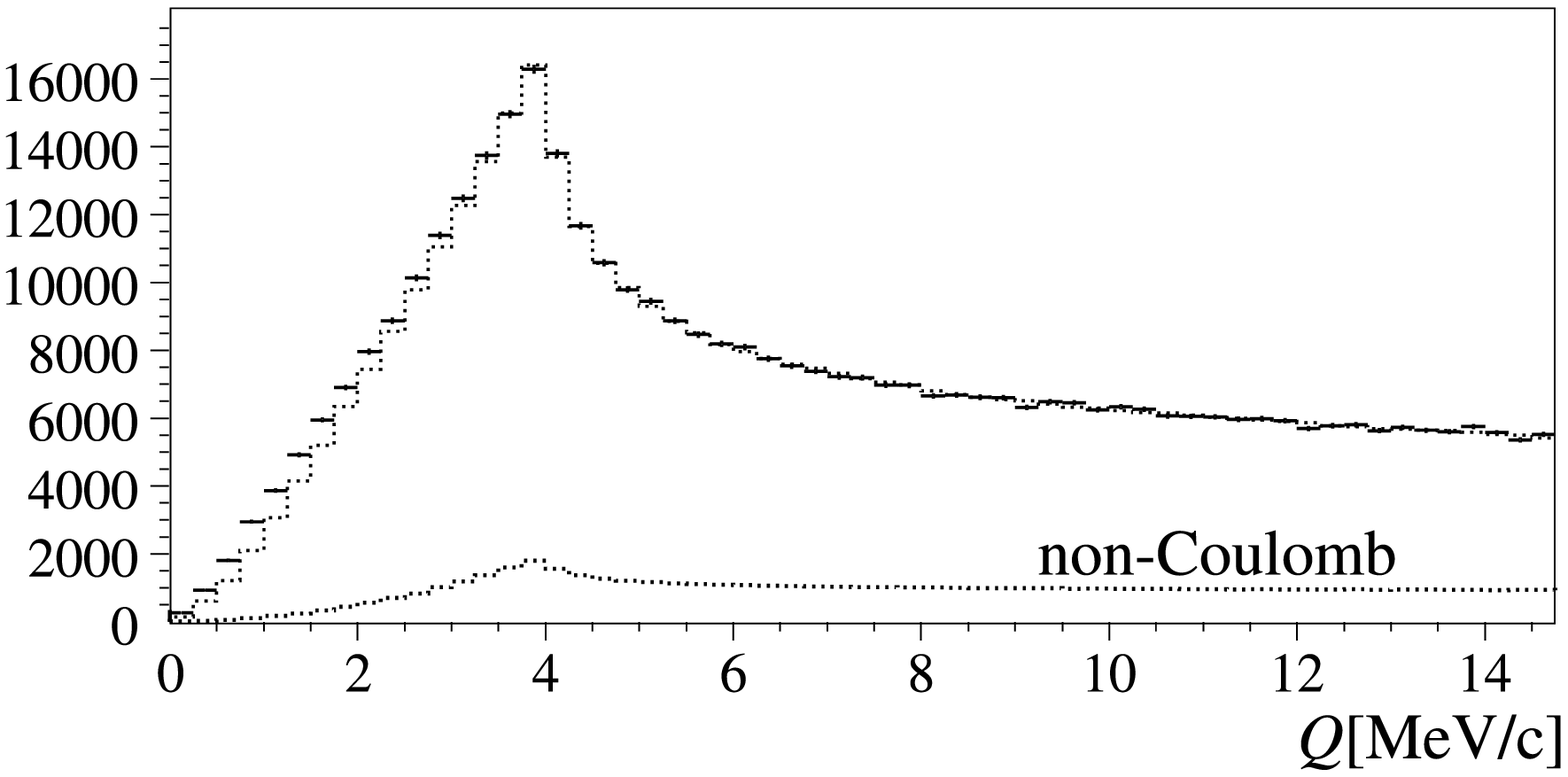}&\includegraphics[width=0.47\textwidth]{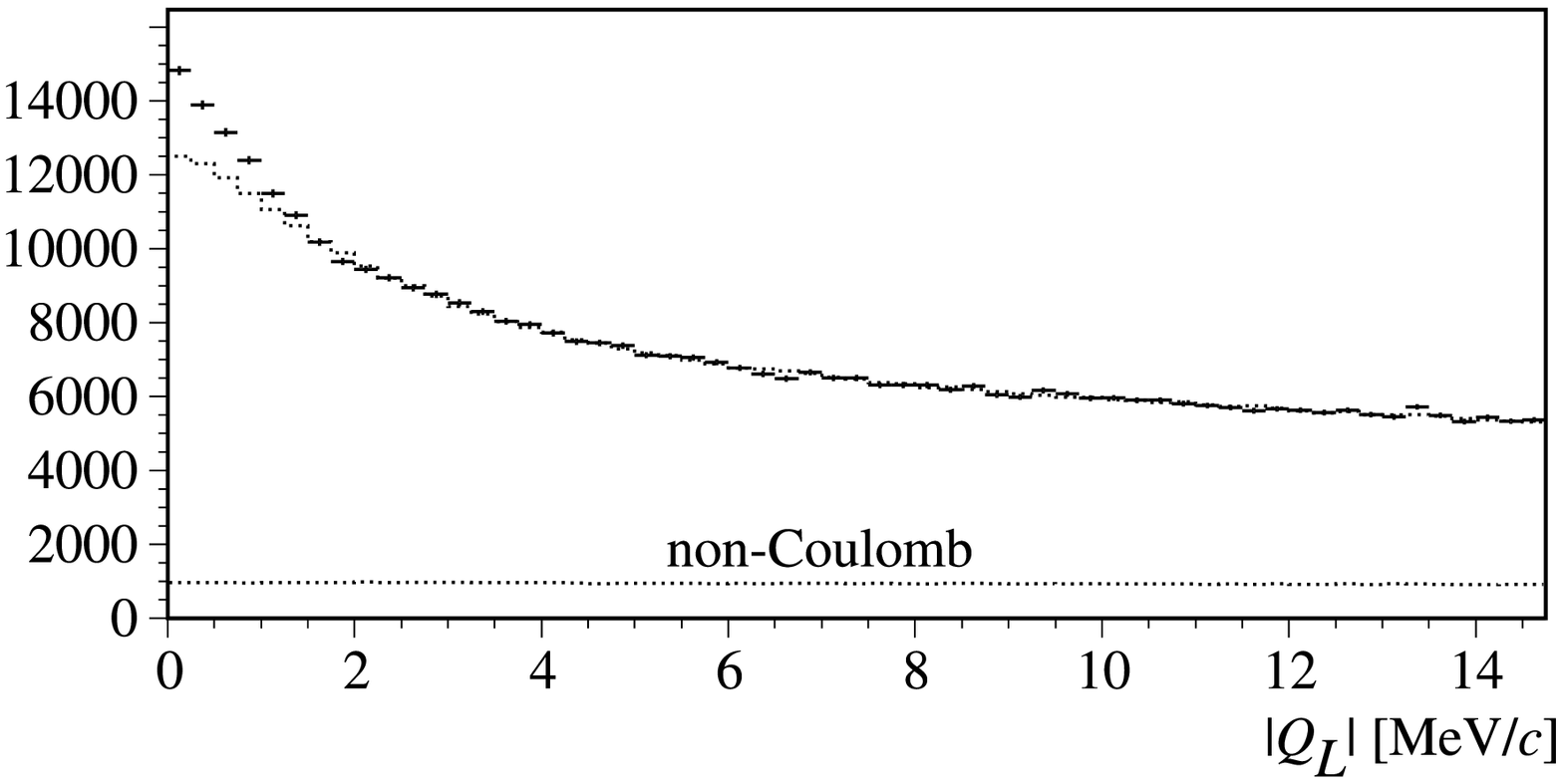}  
   \\   
   \includegraphics[width=0.47\textwidth]{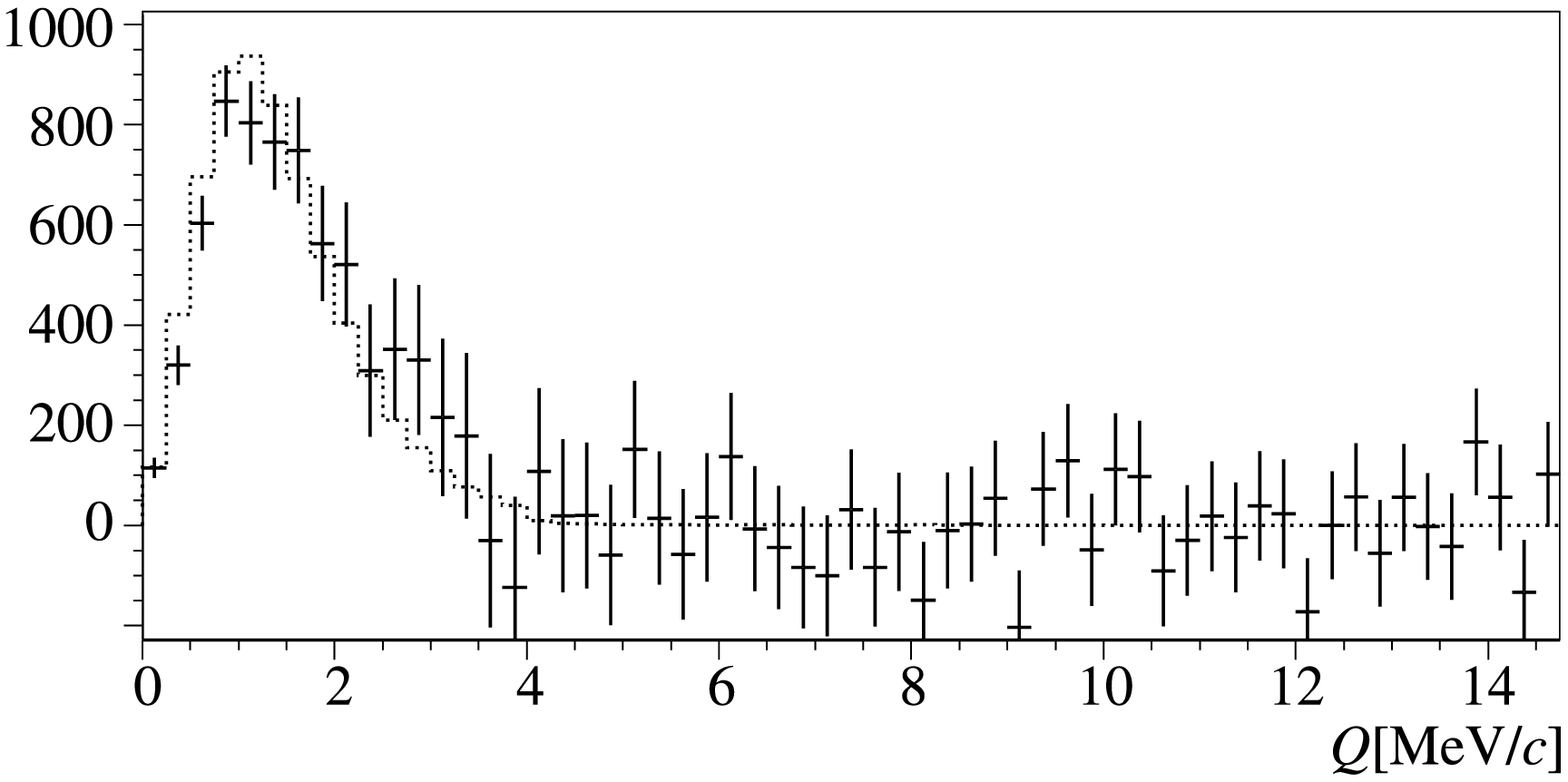}&\includegraphics[width=0.47\textwidth]{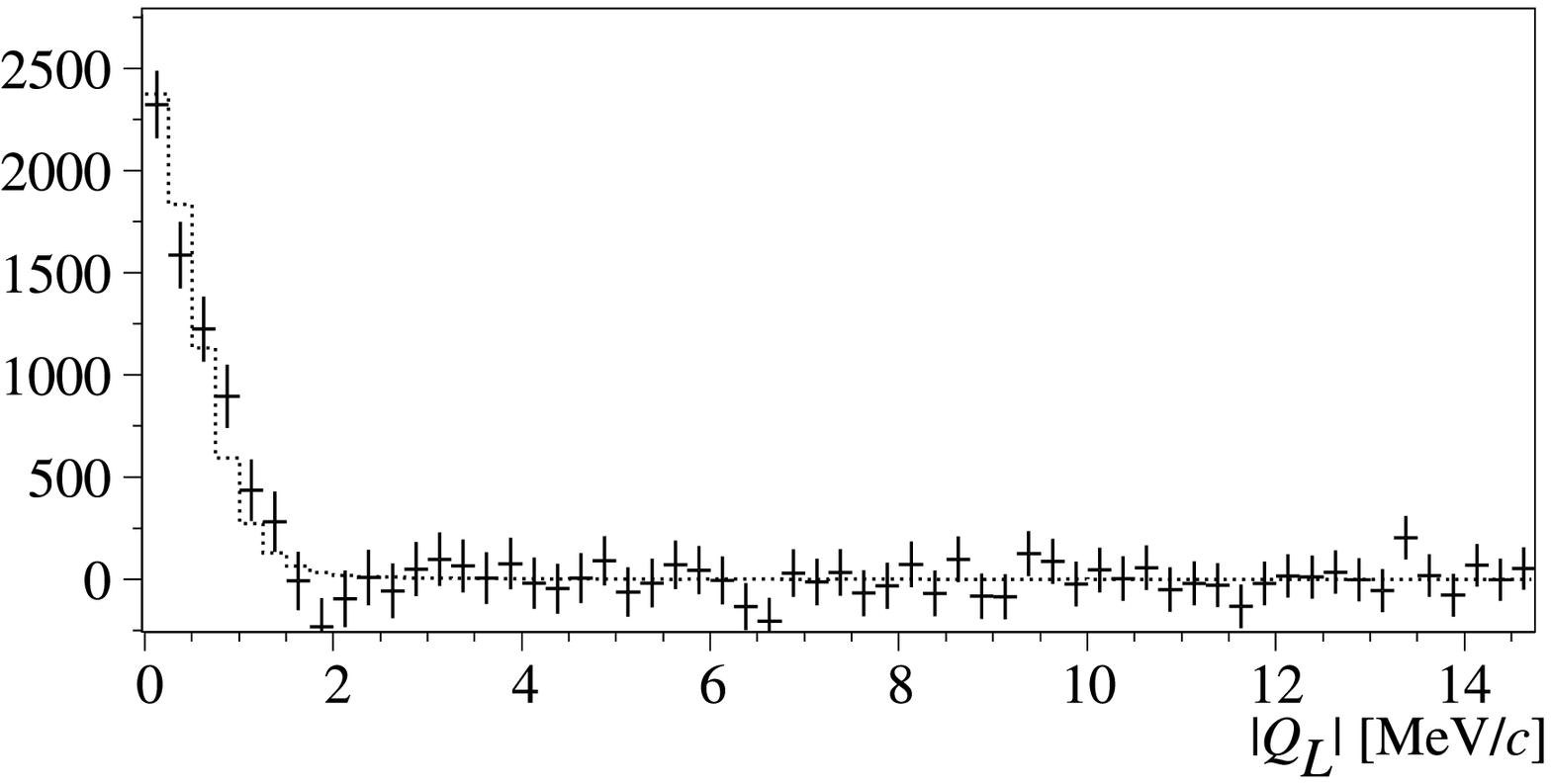}
\end {array}$
\caption 
{
Top: Experimental $Q$ and $Q_{L}$ 
distributions after subtraction of the prompt accidental background,
and fitted Monte Carlo 
backgrounds (dotted lines). The peak at $Q=4$\,MeV/$c$ is due to the cut
$Q_{T}\leq4$\,MeV/$c$.
 Bottom: Residuals after background subtraction. 
 The dotted lines represent the expected atomic signal shape.
The bin-width is 0.25~MeV/$c$.  }
\label{fig:BS-fit}
\end{figure}

The measured prompt distributions are approximated by appropriate
shape functions. The functions for atomic pairs, $F_{A}(Q)$, and for
the backgrounds, $F_{B}(Q)$, (analogously for $Q_{L}$) are defined as:

\begin{equation}
    \centering
\begin{array}{lll}
    F_{A}(Q)&=&\displaystyle{\frac{n_{A}^{rec}}{n_{A}^{MC}}\frac{dn_{A}^{MC}}{dQ}  }\\
    F_{B}(Q)&=&\displaystyle{\frac{N_{CC}^{rec}}{N_{CC}^{MC}}\frac{dN_{CC}^{MC}}{dQ}
    +\frac{N_{NC}^{rec}}{N_{NC}^{MC}}\frac{dN_{NC}^{MC}}{dQ}
    +\frac{\omega_{acc}N_{pr}}{N_{acc}^{MC}}\frac{dN_{acc}^{MC}}{dQ}
    }
\end{array}
\label{eq:fitfunc}
\end{equation}

with $n_{A}^{rec},~N_{CC}^{rec},~N_{NC}^{rec}$ the reconstructed
number of atomic pairs, Coulomb- and non-correlated background,
respectively, and $\omega_{acc}$ the fraction of accidental background
out of all prompt events $N_{pr}$. Analyzing the time distribution
measured with the vertical hodoscopes (see \cite{signalpaper}) we find
$\omega_{acc}$=7.1\% (7.7\%) for the 94 $\mu$m (98 $\mu$m) data sets
\cite{signalpaper,schuetzthesis} and keep it fixed when fitting.  The
$\chi^{2}$ function for $Q$ (analogously for $Q_{L}$) to minimize is

\begin{equation}
    \chi^{2}=  \sum_{\nu_{min}}^{\nu_{max}}
    \frac{\left[\left(\frac{dN_{pr}}{dQ}\Delta Q\right)_{\nu}  
    -\left(\left[F_{A}(Q)+F_{B}(Q)\right]\Delta Q\right)_{\nu}\right]^{2}}
    {\left(\frac{dN_{pr}}{dQ}\Delta Q\right)_{\nu}
    +(\sigma_{A})_{\nu}^{2}
    +(\sigma_{B})_{\nu}^{2}}
    \label{eq:chisq}
\end{equation}

with $\Delta Q$ the bin width and $\sigma_{A}$, $\sigma_{B}$ the
statistical errors of the Monte Carlo shape functions, which are much
smaller than that of the measurement. The fit parameters are
$n_{A}^{rec},~N_{CC}^{rec},~N_{NC}^{rec}$ (see Eq.
(\ref{eq:fitfunc})). As a constraint the total number of measured
prompt events is restricted by the condition
$N_{pr}~(1-\omega_{acc})=N_{CC}^{rec}+N_{NC}^{rec}+n_{A}^{rec}$.  The
measured distributions as well as the background are shown in Fig.
\ref{fig:BS-fit} (top).

The data taken with 94 and 98 $\mu$m thick targets were analyzed
separately. The total number of events in the prompt window is
$N_{pr}=471290$.

First, we determine the background composition by minimizing Eq.
(\ref{eq:chisq}) outside of the atomic pair signal region, i.e. for
$Q> 4 \mathrm{MeV}/c$ and $Q_{L}> 2 \mathrm{MeV}/c$. For this purpose
we require $n_{A}^{rec}=0$. As a constraint, the background parameters
$N_{CC}^{rec}$ and $N_{NC}^{rec}$ representing the total number of
$CC$- and $NC$-events, have to be the same for $Q$ and $Q_{L}$. Then,
with the parameters found, the background is subtracted from the
measured prompt distribution, resulting in the residual spectra.  For
the signal region, defined by the cuts $Q=4 \mathrm{MeV}/c$ and
$Q_{L}= 2 \mathrm{MeV}/c$, we obtain the total number of atomic pairs,
$n_{A}^{residual}$ and of Coulomb correlated background events,
$N_{CC}^{sig}$. Results of fits for $Q$ and $Q_{L}$ together are shown
in Table \ref{tbl:results}.

\begin{table}[tbp] 
    \centering
    \caption{Fit results (94 and 98 $\mu$m targets together, 
    background shapes from Monte Carlo (MC)) for the parameters $N_{CC}^{rec}$
    (total number of CC-events), $N_{NC}^{rec}$ (total number of NC
    events) and $n_{A}^{rec}$ (atomic pairs) and deduced
    results for the number of atomic pairs from the residuals ($n_{A}^{residual}$)
    and the number of CC-background events in the
    signal region ($N_{CC}^{sig}$). MC-a: background fit excluding
    the signal region. MC-b:
    fit of the entire momentum range including Monte Carlo shape for atomic
    pairs (``shape fit''). The cuts were at  $Q_{cut}=4 \mathrm{MeV}/c$ and $Q_{L,~cut}= 2 \mathrm{MeV}/c$.
    $Q$ and $Q_{L}$-distributions were fitted together. The normalized
    $\chi^{2}$ were 0.9 for MC-a and MC-b.}\vspace{0.3cm}
 \begin{tabular}{|c|c|ccccc|} \hline
    & & $N_{CC}^{rec}$     & $N_{NC}^{rec}$          
    &$n_{A}^{residual}$&$n_{A}^{rec}$&$N_{CC}^{sig}$  \\ \hline
    MC-a&Q &$374022\pm 3969$&$56538 $&$6518 \pm 373$&&$106500\pm 1130$ \\
    &$Q_{L}$& same         & same & $6509 \pm 330$ & & $82289 \pm 873$ \\ \hline    
    MC-b&Q &$374282\pm 3561$&$56213$& &$6530 \pm 294$&$106549 \pm 1014$ \\
    &$Q_{L}$&  same       &   same               &   &  same          & $82345 \pm
    783$ \\ \hline
  \end{tabular}
     \label{tbl:results}
\end{table}

CC-background and NC- or acc-backgrounds are distinguishable due to
their different shapes, most pronounced in the $Q_{L}$ distributions
(see Fig. \ref{fig:BS-fit}, top). Accidental and NC-background shapes
are almost identical for $Q$ and fully identical for $Q_{L}$ (uniform
distributions). Thus, the errors in determining the accidental
background $\omega_{acc}$ are absorbed in fitting the NC background.
The correlation coefficient between CC and NC background is $-99\%$.
This strong correlation leads to equal errors for $N_{CC}^{rec}$ and
$N_{NC}^{rec}$. The CC-background is determined with a precision
better than 1\%.  Note that the difference between all prompt events
and the background is
$N_{pr}-N_{CC}^{rec}-N_{NC}^{rec}-\omega_{acc}N_{pr}=6590$, hence very
close to the number of residual atomic pairs ($n_{A}^{residual}$) as
expected. This relation is also used as a strict constraint for fits
outside of the signal region ($>$),
$N_{pr}^{>}-N_{CC}^{rec>}-N_{NC}^{rec>} -(\omega_{acc}N_{pr})^{>}=0$
and, hence, the fit requires only one free parameter, $N_{CC}^{rec>}$.

Second, the atomic pair signal may be directly obtained by minimizing
Eq.  (\ref{eq:chisq}) over the full range and including the Monte
Carlo shape distribution $F_{A}$ (``shape fit''). The signal strength
has to be the same for $Q$ and $Q_{L}$. The result for the signal
strength $n_{A}^{rec}$ as well as the CC-background below the cuts,
$N_{CC}^{sig}$, are shown in Table \ref{tbl:results}.  The errors are
determined by MINOS \cite{minuit}.

The consistency between the analysis in $Q$ with the one in $Q_{L}$
establishes the correctness of the $Q_{T}$ reconstruction. A 2D fit in
the variables ($Q_{L},~Q_{T}$) 
confirms the results of Table \ref{tbl:results}.


\section{Break-up probability}

In order to deduce the break-up probability, $P_{br}=n_{A}/N_{A}$, the
total number of atomic pairs $n_{A}$ and the total number of produced
$A_{2\pi}$ atoms, $N_{A}$, have to be known. None of the two numbers
is directly measured. The procedure of obtaining the two quantities
requires reconstruction efficiencies and is as follows.

\textbf{Number of atomic pairs:} Using the generator for atomic pairs
a large number of events, $n_{A}^{gen}$, is generated in a predefined
large spatial acceptance window $\Omega_{gen}$, propagated through
GEANT-DIRAC including the target and reconstructed along the standard
procedures. The total number of reconstructed Monte Carlo atomic pairs
below an arbitrary cut in $Q$, $n_{A}^{MC-rec}(Q\le Q_{cut})$ defines
the reconstruction efficiency for atomic pairs
$\epsilon_{A}^{cut}=n_{A}^{MC-rec}(Q\le Q_{cut})/n_{A}^{gen}$. The
total number of atomic pairs is obtained from the measured pairs by $
n_{A}=n_{A}^{rec}(Q\le Q_{cut})/\epsilon_{A}^{cut}$.

\textbf{Number of produced $A_{2\pi}$ atoms:} Here we use the known
relation between produced atoms and Coulomb correlated
$\pi^{+}\pi^{-}$ pairs (CC-background) of Eq. (\ref{eq:kfactor}).
Using the generator for CC pairs, $N_{CC}^{gen}$ events, of which
$N_{CC}^{gen}(q\le q_{0})$ (see Eq. \ref{eq:kth}) have $q$ below
$q_{0}$, are generated into the same acceptance window $\Omega_{gen}$
as for atomic pairs and processed analogously to the paragraph above
to provide the number of reconstructed CC-events below the same
arbitrary cut in $Q$ as for atomic pairs, $N_{CC}^{MC-rec}(Q\le
Q_{cut})$.  These CC-events are related to the originally generated
CC-events below $q_{0}$ through
$\epsilon_{CC}^{cut}=N_{CC}^{MC-rec}(Q\le Q_{cut})/N_{CC}^{gen}(q\le
q_{0})$.  The number of produced atoms thus is
$N_{A}=k_{th}(q_{0})N_{CC}^{rec}(Q\le Q_{cut})/\epsilon_{CC}^{cut}$
(see Eq. (\ref{eq:kth})).

The break-up probability $P_{br}$ thus becomes:

\begin{equation}
    P_{br}=\frac{n_{A}}{N_{A}}=
    \frac{n_{A}^{rec}(Q\le Q_{cut})}{k(Q_{cut})N_{CC}^{rec}(Q\le Q_{cut})}
    \qquad\mathrm{with}\;\;
    k(Q_{cut})=k_{th}(q_{0})\frac{\epsilon_{A}^{cut}}{\epsilon_{CC}^{cut}}.
    \label{eq:Pbr}
\end{equation}

In Table \ref{tbl:kfactors} the $k$-factors are listed for different
cuts in $Q$ and $Q_{L}$ for the two target thicknesses (94$\mu$m and
98$\mu$m) and the weighted average of the two, corresponding to their
relative abundances in the Ni data of 2001. The accuracy is of the
order of one part per thousand and is due to Monte Carlo statistics.

\begin{table}[tbp]
    \centering
    \caption{$k(Q_{cut})$ factors
    as a function of cuts in $Q$ and $Q_{L}$ for the
    94  and 98 $\mu$m thick Ni targets, and the weighted average of
    the two for a relative abundance of 76\% (94$\mu$m) and 24\%
    (98$\mu$m).}\vspace{0.3cm}
     \begin{tabular}{|l|ccc|} \hline
   & $k_{94~\mu\mathrm{m}}$ & $k_{98~\mu\mathrm{m}}$  & $k_{average}$\\  \hline
  $Q_{cut}=2\mathrm{MeV}/c$ &$0.5535\pm 0.0007$&$0.5478\pm0.0007$&$0.5521\pm 0.0007$ \\  \hline
  $Q_{cut}=3\mathrm{MeV}/c$ &$0.2565\pm 0.0003$&$0.2556\pm0.0003$&$0.2563\pm 0.0003$ \\  \hline
  $Q_{cut}=4\mathrm{MeV}/c$ &$0.1384\pm 0.0002$&$0.1383\pm0.0002$&$0.1384\pm 0.0002$ \\  \hline
  $Q_{L,cut}=1\mathrm{MeV}/c$ &$0.3054\pm 0.0004$&$0.3044\pm0.0003$&$0.3050\pm 0.0004$ \\  \hline
  $Q_{L,cut}=2\mathrm{MeV}/c$ &$0.1774\pm  0.0002$&$0.1776\pm0.0002$&$0.1774\pm 0.0002$ \\  \hline  
    \end{tabular} 
     \label{tbl:kfactors}
\end{table}

With the $k$-factors of Table \ref{tbl:kfactors} and the measurements
listed in Table \ref{tbl:results}, the break-up probabilities of Table
\ref{tbl:PBR} are obtained. The simultaneous fit of $Q$ and $Q_{L}$
with the atomic shape results in a single value.

The break-up probabilities from $Q$ and $Q_{L}$ agree within a
fraction of a percent. The values from shape fit and from background
fit are in perfect agreement (see Table \ref{tbl:results}). We adopt
the atomic shape fit value of $P_{br}=0.447\pm 0.023_{stat}$, because
the fit covers the full $Q$, $Q_{L}$ range and includes correlations
between $n_{A}^{rec}$ and $N_{CC}^{sig}$.
 
Analyzing the data with three allowed hit candidates in the SFD search
window instead of two, results in more atomic pairs (see Ref.
\cite{signalpaper}, T-tracking).  The break-up probabilities obtained
are $0.440\pm 0.024$ and $0.430\pm 0.021$ for $Q$ and $Q_{L}$,
respectively. They are not in disagreement with the adopted value of
0.447. Despite the larger statistics, the accuracy is not improved,
due to additional background.  This background originates from
additional real hits in the upstream detectors or from electronic
noise and cross-talk. This has been simulated and leads essentially to
a reduced reconstruction efficiency but not to a deterioration of the
reconstruction quality. The additional sources of systematic
uncertainties lead us not to consider this strategy of analysis
further on.

V-tracking provides a slightly different data sample, different
$k$-factors and different signal strengths and CC-background.  The
break-up probability, however, does not change significantly and is
$P_{br}^{V-tracking}=0.453\pm 0.025_{stat}$, only 0.3 $\sigma$ off
from the adopted value 0.447.

The break-up probability has to be corrected for the impurities of the
targets. Thus, the 94 $\mu$m thick target has a purity of only 98.4\%,
while the 98 $\mu$m thick target is 99.98 \% pure. The impurities (C,
Mg, Si, S, Fe, Cu) being mostly of smaller atomic number than Ni lead
(for the weighted average of both targets) to a reduction of the
break-up probability of 1.1\% as compared to pure Ni, assuming a
lifetime of 3$fs$.  Therefore, the measured break-up probability has
to be increased by 0.005 in order to correspond to pure Ni. The final
result is:

\begin{equation}
    P_{br}=0.452\pm 0.023_{stat}.
    \label{eq:Pbr-result}
\end{equation}

\begin{table}[tbp] 
    \centering
    \caption{Break-up probabilities for the combined Ni2001 data,
    based on the results of Table \ref{tbl:results} and the
    $k$-factors of Table \ref{tbl:kfactors} for the cuts $Q_{cut}=4\mathrm{MeV}/c$ 
    and $Q_{L,cut}=2\mathrm{MeV}/c$ . Errors are
    statistical.}\vspace{0.3cm}
 \begin{tabular}{|c|cccc|} \hline
     & $n_{A}^{residual}$    & $n_{A}^{rec}$&$N_{CC}^{sig}$&$P_{br}$ \\ \hline
    $Q$       &$6518 \pm 373$&&$ 106500\pm 1130$   &$0.442\pm 0.026$\\
    $Q_{L}$&$6509 \pm 330$&& $82289\pm 873$ &$0.445\pm 0.023$\\ \hline
    $Q$ \&$Q_{L}$ &  &$6530 \pm 294$&$106549 \pm 1004$&$0.447\pm 0.023$ \\ \hline
  \end{tabular}
     \label{tbl:PBR}
\end{table}


\section{Systematic errors}

Systematic errors may occur through the analysis procedures and
through physical processes which are not perfectly under control. We
investigate first procedure-induced errors.

\begin{figure}[htb]
\begin{minipage}{0.45\textwidth}
\centering 
    \includegraphics[width=0.90\textwidth]{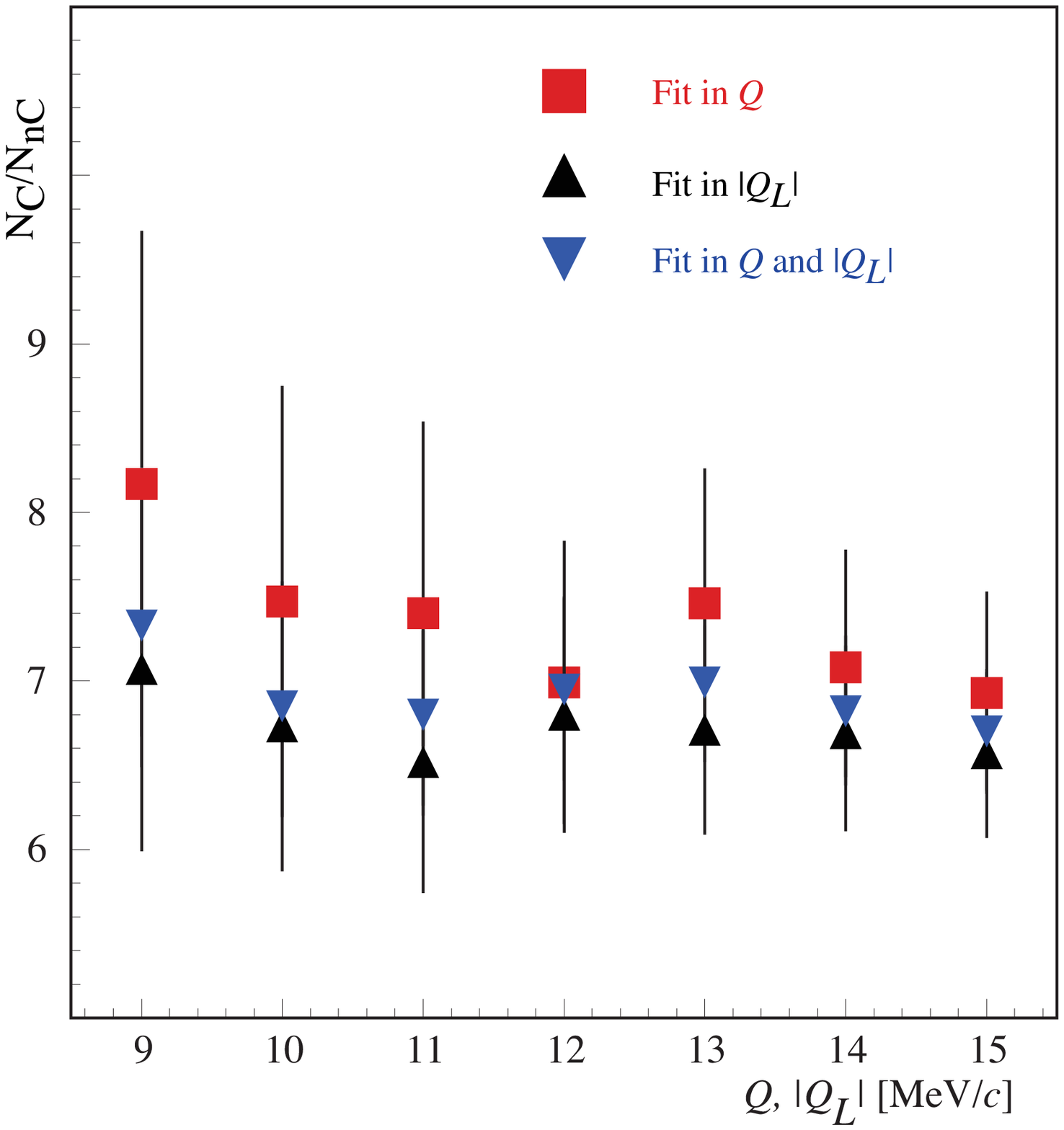}
    \caption{Ratio of CC-background over NC-background as a function of 
    fit range.}
    \label{fig:pbrrange}
\end{minipage}
\hfill
\begin{minipage}{0.45\textwidth}
\centering
\centering
     \includegraphics[width=0.90\textwidth]{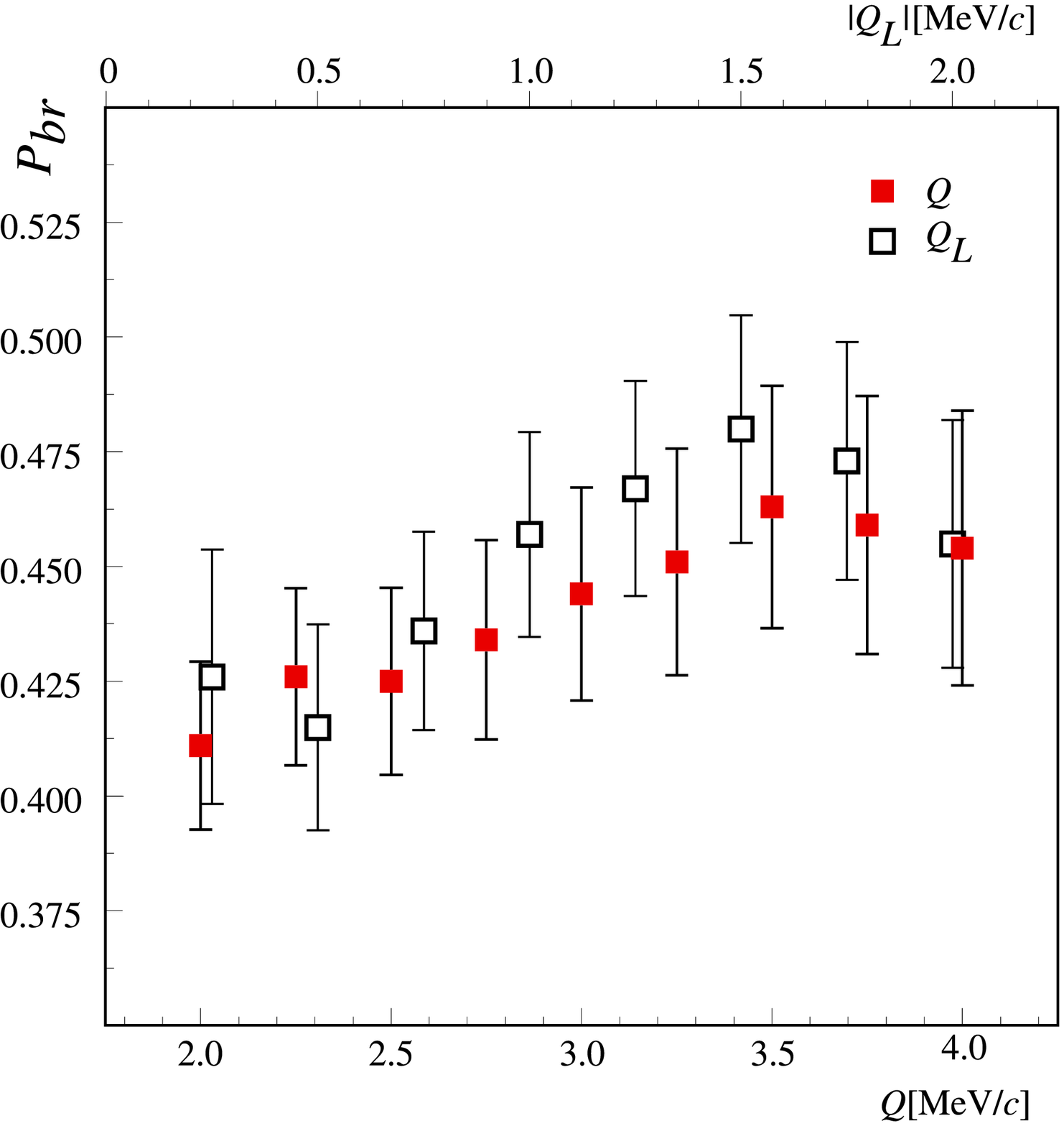}
    \caption{$P_{br}$ as a function of cut momentum for $Q$ and $Q_{L}$.}
    \label{fig: Pbrvscut}
\end{minipage}
\end{figure}

The break-up probability will change, if the ratio
$N_{CC}^{rec}/N_{NC}^{rec}$ depends on the fit range.  If so, the
Monte Carlo distributions do not properly reproduce the measured
distributions and the amount of CC-background may not be constant. In
Fig. \ref{fig:pbrrange} the dependence is shown for the fits in $Q$,
$Q_{L}$ and both together. The ratio is reasonably constant within
errors, with the smallest errors for a fit range of $Q=Q_{L}=15
\mathrm{MeV}/c$.  At this point the difference between $Q$ and $Q_{L}$
fits leads to a difference in break-up probability of $\Delta
P_{br}^{CC}=0.023$.

Consistency of the procedure requires that the break-up probability
does not depend on $Q_{cut}$. In Figure \ref{fig: Pbrvscut} the
dependence on the cut is shown for break-up probabilities deduced from
$n_{A}^{residual}$. There is a systematic effect which, however,
levels off for large cut momenta. This dependence indicates that the
shape of the atomic pair signal as obtained from Monte Carlo (and used
for the $k$-factor determination) is not in perfect agreement with the
residual shape. This may be due to systematics in the atomic pair
shape directly and/or in reconstructed CC-background for small
relative momenta. The more the signal is contained in the cut, the
more the $P_{br}$ values stabilize. As a consequence, we chose a cut
that contains the full signal (see Eq. (\ref{eq:Pbr-result})). This
argument is also true for sharper cuts in $Q_{T}$ than the one from
the event selection. Cut momenta beyond the maximum cut of Figure  \ref{fig: Pbrvscut}
would only test background, as the signal would not change anymore.
 
To investigate whether the
atomic pair signal shape is the cause of the above cut dependence, we 
studied two
extreme models for atom break-up: break-up only from
the 1$S$-state and break-up only from highly excited states. The two extremes result
in a difference in break-up probability of $\Delta
P_{br}^{shape}=0.008$.
 
Sources of systematic errors may also arise from uncertainties in the 
genuine
physical process. We have investigated possible uncertainties in
multiple scattering as simulated by GEANT by changing the scattering
angle in the GEANT simulation by $\pm 5\% $. As a result, the break-up
probability changes by 0.002 per one percent change of multiple
scattering angle. In fact we have measured the multiple scattering for
all scatterers (upstream detectors, vacuum windows, target) and found
narrower angular distributions than expected from the standard GEANT
model \cite{kruglov}. This, however, may be due also to errors in
determining the thickness and material composition of the upstream
detectors. Based on these studies we conservatively attribute a
maximum error of +5\% and -10\% to multiple scattering.

Another source of uncertainty may be due to the presence of
unrecognized $K^{+}K^{-}$ and $\bar{p} p$ pairs that would fulfill all
selection criteria \cite{KKpp}. Such pairs may be as abundant as 0.5\%
and 0.15\%, respectively, of $\pi^{+}\pi^{-}$ pairs as estimated for
$K^{+}K^{-}$ with FRITIOF-6 \footnote{FRITIOF-6 reproduces well
production cross sections and momentum distributions for 24\,GeV/$c$
proton interactions.}  and for $\bar{p} p$ from time-of-flight
measurements in a narrow momentum interval with DIRAC data. Their mass
renders the Coulomb correlation much more peaked at low $Q$ than for
pions, which leads to a change in effective $\pi^{+}\pi^{-}$ Coulomb
background at small $Q$, thus to a smaller atomic pair signal and
therefore to a decrease of break-up probability.  The effect leads to
a change of $\Delta P_{br}^{\bar{K}K,~\bar{p} p}=-0.04$. We do not
apply this shift but consider it as a maximum systematic error of
$P_{br}$. Admixtures from unrecognized $e^{+}e^{-}$ pairs from photon
conversion do not contribute because of their different shapes.

Finally, the correlation function Eq. (\ref{eq:coulomb}) used in the
analysis is valid for pointlike production of pions, correlated only
by the Coulomb final state interaction (Eq. (\ref{eq:ac})). However, there
are corrections due to finite size and strong interaction
\cite{lednicki}.  These have been studied based on the UrQMD transport
code simulations \cite{transportcode} and DIRAC data on
$\pi^{-}\pi^{-}$ correlations.  The parameters of the underlying model
are statistically fixed with data up to 200\,MeV/$c$ relative momentum.
For $Q\leq 30$\,MeV/$c$, the DIRAC data are too scarce to serve as a
test of the model. The corrections lead to a change of $\Delta
P_{br}^{finite-size}=-0.02$.  Due to the uncertainties we
conservatively consider 1.5 times this change as a maximum error, but
do not modify $P_{br}$.

The systematics are summarized in Table \ref{tbl:syst}.  The extreme
values represent the ranges of the assumed uniform probability density
function (u.p.d.f.), which, in case of asymmetric errors, were
complemented symmetrically for deducing the corresponding standard
deviations $\sigma $. Convoluting the five u.p.d.f. results in
bell-shaped curves very close to a Gaussian, and the $\pm \sigma$
(Table \ref{tbl:syst}, total error) correspond roughly to a 68.5\%
confidence level and can be added in quadrature to the statistical
error.

The final value of the break-up probability is 

\begin{equation}
    P_{br}=0.452\pm
    0.023_{stat}~^{+0.009}_{-0.032}\}_{syst}=0.452~^{+0.025}_{-0.039}.
    \label{eq:finalpbr}
\end{equation}

\begin{table}[tbp]
    \centering
    \caption{Summary of systematic effects on the measurement of the
    break-up probability $P_{br}$. Extreme values have been
    transformed into $\sigma $ assuming uniform distributions.
    }\vspace{0.3cm}
    \begin{tabular}{|c|c|c|}
        \hline
        source & extreme values & $\sigma$  
        \\ \hline
        CC-background & +0.012 / -0.012& $\pm$0.007  
       \\ \hline
        signal shape & +0.004 / -0.004& $\pm$0.002  
        \\ \hline
        multiple scattering & +0.01~/-0.02 &$~^{+0.006}_{-0.013}$
        \\ \hline
        $K^{+}K^{-}$ and $\bar{p}p$ &$ +0 ~/ -0.04$ &$~^{+0}_{-0.023}$ 
        \\ \hline
        finite size &$ +0 ~/ -0.03$ &$  ~^{+0}_{-0.017}$  
        \\ \hline \hline 
        Total &$     $.&$ ~^{+0.009}_{-0.032}$ 
        \\ \hline       
    \end{tabular}
    
    \label{tbl:syst}
\end{table}


\section{Lifetime of Pionium}

\begin{figure}[htb]
\centering
  \includegraphics[width=0.50\textwidth]{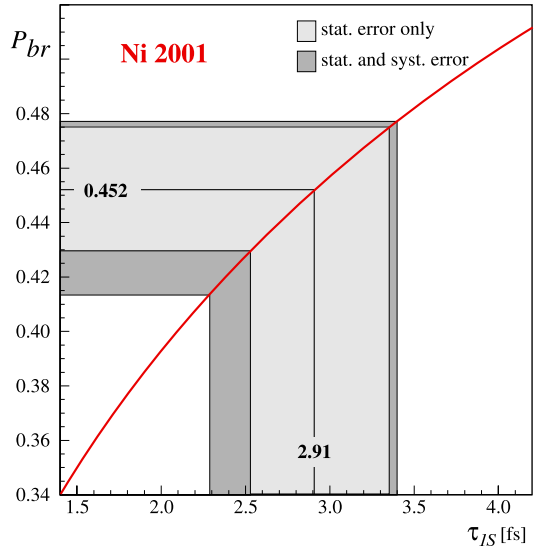}
\caption{ Break-up probability $P_{br}$ as a function of the lifetime 
  of the atomic ground state $\tau_{1S}$ for the combined 94 and 98
  $\mu$m thick Ni targets. The experimentally determined $P_{br}$ with
  statistical and total errors translates into a value of the lifetime
  with corresponding errors.}
\label{fig:tauresult}
\end{figure}

The lifetime may be deduced on the basis of the relation between
break-up probability and lifetime for a pure Ni target (Fig.
\ref{fig:tauresult}). This relation, estimated to be accurate at the
1\% level, may itself have uncertainties due to the experimental
conditions. Thus the target thickness is estimated to be correct to
better than $\pm~1~\mu$m, which leads to an error in the lifetime (for
$P_{br}=0.45$) smaller than $\pm ~0.01$ fs, less than 1\% of the
expected lifetime and thus negligible. The result for the lifetime is
\begin{equation}
    \tau_{1S}=\left[2.91~^{+0.45}_{-0.38}\}_{stat}~^{+0.19}_{-0.49}\}_{syst}\right]\times
    10^{-15} ~\mathrm{s}=\left[2.91~^{+0.49}_{-0.62}\right]\times
    10^{-15} ~\mathrm{s}.
    \label{eq:tauresult}
\end{equation}

The errors are not symmetric because the $P_{br}-\tau $ relation is
not linear, and because finite size corrections and heavy particle
admixtures lead to possible smaller values of $P_{br}$. The accuracy
achieved for the lifetime is about $+17\%$, almost entirely due to
statistics and $-21\%$, due to statistics and systematics in roughly
equal parts. With full statistics (2.3 times more than analysed here)
the statistical errors may be reduced accordingly. The two main
systematic errors (particle admixtures and finite size correction)
will be studied in more detail in the future program of DIRAC.

Using Eq. (\ref{eq:gasser}), the above lifetime corresponds to
$\left|a_0-a_2\right|=0.264~^{+0.033}_{-0.020}~m_{\pi}^{-1}$.

\section{Acknowledgements}
We are indebted to the CERN PS crew for providing a beam of excellent quality. 
This work was supported by CERN,
the Grant Agency of the Czech Republic, grant No. 202/01/0779 and
202/04/0793, 
the Greek General Secretariat of Research and Technology (Greece), 
the University of Ioannina Research Committee (Greece), 
the IN2P3 (France), 
the Istituto Nazionale di Fisica Nucleare (Italy),
the Grant-in-Aid for Scientific Research from Japan Society for the Promotion of Science 07454056, 
08044098, 09640376, 09440012, 11440082, 11640293, 11694099, 12440069, 14340079, and
15340205,
the Ministry of Education and Research, under project CORINT No.1/2004
(Romania),
the Ministery of Industry, Science and Technologies of the Russian
Federation and the Russian Foundation for Basic Research (Russia),
under project 01-02-17756, the Swiss National Science Foundation, the
Ministerio de Ciencia y Tecnologia (Spain), under projects AEN96-1671
and AEN99-0488, the PGIDT of Xunta de Galicia (Spain).


\begin{thebibliography}{99}
\bibitem{Adeva95} B. Adeva et al., DIRAC proposal, CERN/SPSLC 95-1,
        SPSLC/P 284 (1995).
\bibitem{Uretsky61} J.~Uretsky and J.~Palfrey, Phys. Rev. 121 (1961) 1798.

\bibitem{hammer} H.-W. Hammer and J.N. Ng, Eur. Phys.J. A6 (1999) 115.

\bibitem{Deser54} S.~Deser et al., Phys. Rev. 96 (1954) 774.

\bibitem{Bilenky69}  S.M.~Bilenky et al., Yad. Phys. 10 (1969) 812;
        (Sov.\ J.\ Nucl.\ Phys. 10 (1969) 469).
\bibitem{Jallouli98} H.~Jallouli and H.~Sazdjian, Phys. Rev. D58 (1998)
        014011; Erratum: ibid., D58 (1998) 099901.
\bibitem{Ivanov98} M.A.~Ivanov et al. Phys. Rev. D58 (1998) 094024.

\bibitem{Gasser01} J.~Gasser et al., Phys. Rev. D64 (2001) 016008;
        hep-ph/0103157.

\bibitem{Gashi02} A.~Gashi et al., Nucl. Phys. A699 (2002) 732.

\bibitem{Weinb79} S.~Weinberg, Physica A96 (1979) 327; J.~Gasser and
        H.~Leutwyler, Phys. Lett. B125 (1983) 325; ibid Nucl. Phys. B250
        465, 517, 539.
\bibitem{Colan01NP} G. Colangelo, J. Gasser and H. Leutwyler, Nucl.
        Phys. B603 (2001) 125.

\bibitem{Knecht95} M.~Knecht et al., Nucl. Phys. B457 (1995) 513.

\bibitem{Ross77} L.~Rosselet et al., Phys. Rev. D15 (1977) 547.

\bibitem{Pislak01} S.~Pislak et al., Phys. Rev. Lett, 87 (2001) 221801.

\bibitem{Nem85} L.L.~Nemenov, Yad. Fiz. 41 (1985) 980; (Sov. J. Nucl.
        Phys. 41 (1985) 629).

\bibitem{Sakh48} A.D.Sakharov, Z.Eksp.Teor.Fiz. 18 (1948) 631.

\bibitem{lednicki} R. Lednicky, DIRAC note 2004-06, nucl-th/0501065.

\bibitem{AFAN97} L.G.~Afanasyev, O.O.~Voskresenskaya, V.V.Yazkov,
Communication JINR P1-97-306, Dubna, 1997.


\bibitem{Afan94} L.G.~Afanasyev et al., Phys. Lett. B338 (1994) 478.

\bibitem{Dulian83} L.S.~Dulian and A.M.~Kotsinian, Yad.Fiz. 37 (1983)
        137; (Sov. J. Nucl. Phys. 37 (1983) 78).

\bibitem{Mrowc} S.~Mr\'owczy\'nski, Phys. Rev. A33 (1986) 1549;
        S.Mr\'owczy\'nski, Phys. Rev. D36 (1987) 1520;
        K.G.~Denisenko and S.~Mr\'owczy\'nski, ibid. D36 (1987) 1529.
\bibitem{Afan96} L.G.~Afanasyev and A.V.~Tarasov, Yad.Fiz. 59 (1996)
        2212; (Phys. At. Nucl. 59 (1996) 2130).
\bibitem{Halab99} Z.~Halabuka et al., Nucl.Phys. B554 (1999) 86--102.
\bibitem{Taras91} A.V.~Tarasov and I.U.~Khristova, JINR-P2-91-10, Dubna
        1991.
\bibitem{Voskr98} O.O.~Voskresenskaya, S.R.~Gevorkyan and
        A.V.~Tarasov, Phys. At. Nucl. 61 (1998) 1517.
\bibitem{Afan99} L.~Afanasyev, A.~Tarasov and O.~Voskresenskaya,
        J.Phys. G 25 (1999) B7.
\bibitem{Ivanov99gl} D.Yu.~Ivanov, L.~Szymanowski, Eur.Phys.J. A5
        (1999) 117.
\bibitem{Heim00} T.A.~Heim et al., J. Phys. B33 (2000) 3583.
\bibitem{Heim01} T.A.~Heim et al., J. Phys. B34 (2001) 3763.
\bibitem{Schum02} M.~Schumann et al., J. Phys. B35 (2002) 2683.
\bibitem{Afan02} L.~Afanasyev, A.~Tarasov and O.~Voskresenskaya,
        Phys.Rev.D 65 (2002) 096001; hep-ph/0109208.
\bibitem{santa03} C. Santamarina, M. Schumann, L.G. Afanasyev and T. Heim,
J.Phys.\ B 36 (2003) 4273.

\bibitem{afan04}L. Afanasyev  et al., J. Phys B37 (2004) 4749.

\bibitem{Afan93} L.G.~Afanasyev et al., Phys. Lett. B308 (1993) 200.

\bibitem{Setup03} B. Adeva~B et al., Nucl.Instr.Meth. A515 (2003) 467.

\bibitem{signalpaper} B. Adeva et al. J. Phys. G30 (2004) 1929.

\bibitem{schuetzthesis} Ch. P. Schuetz, ``Measurement of the breakup probability 
of $\pi^{+}\pi^{-}$ atoms in a Nickel target with the
 DIRAC spectrometer'', PhD Thesis, Basel, March 2004, http://cdsweb.cern.ch/searc.py?recid=732756.

 \bibitem{geant} P. Zrelov and V. Yazkov, ``The GEANT-DIRAC Simulation
 Program'', DIRAC note 1998-08,
 http://zrelov.home.cern.ch/zrelov/dirac/montecarlo/instruction/instruct26.html
 
 \bibitem{ariane} D. Drijard, M. Hansroul and V. Yazkov, The DIRAC
 offline user's Guide, and
 /afs/cern.ch/user/d/diracoff/public/offline/docs

\bibitem{santa2003-9} C. Santamarina and Ch. P. Schuetz, DIRAC note 2003-9.

 \bibitem{minuit} F. James and M. Roos, Function Minimization and Error
    Analysis, Minuit, CERN Program Library, D506 MINUIT
    
 \bibitem{kruglov} A. Dudarev et al., DIRAC note 2005-02.

    
 \bibitem{KKpp} O.E. Gortchakov and V.V. Yazkov, DIRAC note 2005-01.
 
\bibitem{transportcode} S. A.  Bass et al.,   Prog. Part. Nucl.Phys. 41 (1998) 225.

\end{thebibliography}
\end{document}